\documentclass[12pt]{article}
\usepackage{color}
\usepackage{fullpage}
\usepackage{amsmath}
\usepackage{amssymb}
\usepackage{amscd}
\usepackage{amsthm}
\usepackage{graphicx}
\usepackage{amssymb}
\usepackage{epstopdf}
\usepackage{cancel}
\usepackage{setspace}
\usepackage{slashed}
\usepackage[hang,small]{caption}
\newcommand{\be}{\begin{equation}}
\newcommand{\ee}{\end{equation}}

\newcommand{\mbb}[1]{\mathbb{#1}}
\newcommand{\mrm}[1]{\mathrm{#1}}
\newcommand{\mcal}[1]{\mathcal{#1}}

\theoremstyle{definition}

\numberwithin{equation}{section}
\begin{document}
%---Title ---------------------------------------------------------------
\begin{titlepage}
\bigskip
\rightline{}

\bigskip\bigskip\bigskip\bigskip
\centerline {\Large \bf {Black Droplets}}
\bigskip\bigskip

\begin{center}
\large Jorge E.~Santos$^{1,2}$ and Benson Way$^{2}$
\\
\bigskip\bigskip
{\small
$^{1}${\em Department of Physics, Stanford University, \\
Stanford, CA 94305-4060, USA }}
\\\bigskip\bigskip
{\small
$^{2}${\em Department of Applied Mathematics and Theoretical Physics, \\
University of Cambridge, Wilberforce Road, \\
Cambridge CB3 0WA, UK}}
\\\bigskip\bigskip
\end{center}
\begin{abstract}
Black droplets and black funnels are gravitational duals to states of a large N, strongly coupled CFT on a fixed black hole background.  We numerically construct black droplets corresponding to a CFT on a Schwarzchild background with finite asymptotic temperature.  We find two branches of such droplet solutions which meet at a turning point.  Our results suggest that the equilibrium black droplet solution does not exist, which would imply that the Hartle-Hawking state in this system is dual to the black funnel constructed in \cite{Santos:2012he}.  We also compute the holographic stress energy tensor and match its asymptotic behaviour to perturbation theory.
\end{abstract}
\end{titlepage}

%\tableofcontents

%===MAIN=================================================================
\onehalfspacing

%---  Introduction -------------------------
\begin{section}{Introduction}
The discovery of Hawking radiation and its associated information paradox has led to a deeper understanding of quantum gravity, and formed a basis for the development of holography and the AdS/CFT correspondence \cite{Maldacena:1997re,Gubser:1998bc,Witten:1998qj}. Recently, there have been many attempts to use holography to further our understanding of Hawking radiation.  In particular, while Hawking radiation is mostly understood for free fields on black hole backgrounds, the authors of \cite{Hubeny:2009ru,Hubeny:2009kz,Hubeny:2009rc} apply AdS/CFT to the study of Hawking radiation when these fields are strongly interacting.

The AdS/CFT correspondence conjectures the equivalence between a large-$N$ gauge theory at strong coupling to a classical theory of gravity in one higher dimension.  The correspondence gives us the freedom to choose a fixed, non-dynamical background spacetime for the gauge theory, which translates to a conformal boundary condition on the gravity side.  For a gauge theory background $\mcal B$ in $D-1$ dimensions, this amounts to solving the $D$-dimensional Einstein's equations with a negative cosmological constant
\be
R_{\mu\nu}=\frac{2\Lambda}{D-2}g_{\mu\nu}\,,\qquad\Lambda=-\frac{(D-1)(D-2)}{2\ell^2}\;,
\ee
with a boundary that is conformal to $\mcal B$. 

For the moment, let us consider the case where $\mcal B$ is an asymptotically flat black hole of size $R$ and temperature $T_{\mrm{BH}}$.  Let's also suppose that far from the black hole, the field theory has a temperature $T_\infty$.  The authors of \cite{Hubeny:2009ru} conjectured two families of solutions that describe the gravity dual.  They argue that in the bulk gravity dual, the thermal state far from the boundary black hole is described in the gravity side by a planar black hole, while the horizon of the boundary black hole must extend into a horizon in the bulk.  These two horizons are either connected, yielding a \emph{black funnel} or disconnected, yielding a \emph{black droplet}. These are illustrated in Fig.~\ref{Fig:dropletfunnel}. 

\begin{figure}[th]
\begin{center}
\includegraphics[width=.42\textwidth]{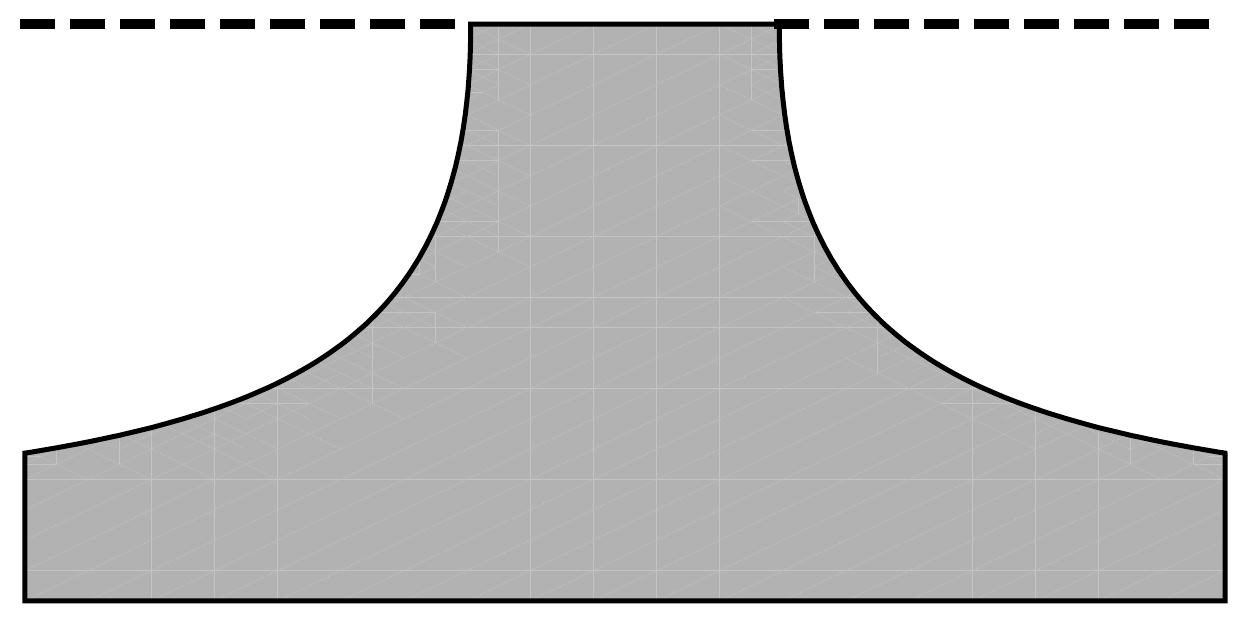}\qquad\qquad
\includegraphics[width=.42\textwidth]{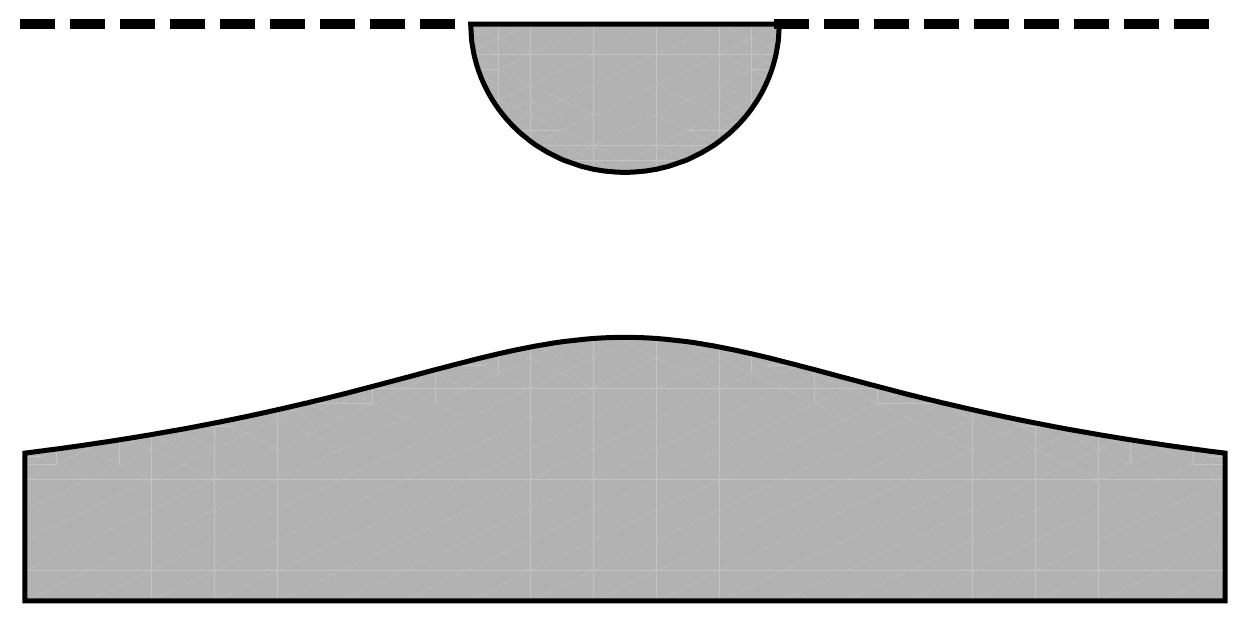}
\end{center}
\caption{Sketches for black funnels (left) and black droplets (right).}\label{Fig:dropletfunnel}
\end{figure}  

In the field theory, the difference between these families is manifest in the way the black hole couples to the thermal bath at infinity.  The connected funnel horizon implies that the field theory black hole readily exchanges heat with infinity.  On the other hand, the disconnected droplet horizons suggest that the coupling between the boundary black hole and the heat bath at infinity is suppressed by $\mcal O(1/N^2)$.  Indeed, unless $T_\mrm{BH}=T_{\infty}$, the funnel solutions would exhibit a ``flowing" geometry\footnote{These flowing funnels would be stationary solutions with non-killing horizons.}.  The droplet solutions, however, are necessarily static for a static boundary black hole.  

A phase transition between these two families would resemble a ``jamming" transition in which a system moves between a more fluid-like phase and a phase with more rigid behaviour.  Based on gravitational intuition for the stability of the bulk solution, it was conjectured in \cite{Hubeny:2009ru} that funnel phases should be preferred for large $RT_\infty$, while droplets should be preferred for small $RT_\infty$.  

In order to test these conjectures, one would need to construct corresponding droplet and funnel solutions.  Droplet solutions are simpler to construct when $T_\infty=0$. In this case, the planar horizon in the droplets becomes the $AdS$ Poincar\'e horizon.  Such droplet solutions were constructed in \cite{Figueras:2011va} for a Schwarzschild boundary, and in \cite{Fischetti:2013hja,Figueras:2013jja} for a boundary that is equal-angular momentum Myers-Perry in 5 dimensions. There is also an analytic droplet based on the C-metric with a three-dimensional boundary black hole \cite{robastefanesei}.  Static funnel solutions (that is, with $T_\mrm{BH}=T_\infty\neq0$) were constructed in \cite{Santos:2012he}, for a Schwarzschild boundary and for a class of 3-dimensional boundary black holes.  

Unfortunately, none of these solutions can be directly compared with each other.  The $T_\infty=0$ droplets will compete with a funnel that flows to zero temperature, and the static funnels compete with a droplet solution with equal temperature horizons.  Neither of these solutions have been constructed.  

In this paper, we shed light on the droplet and funnel transition by numerically constructing new black droplet solutions with $T_\infty\neq0$.  As in \cite{Santos:2012he,Figueras:2011va}, our boundary metric is Schwarzschild. We find that there can be two black droplet solutions for a given $T_\infty/T_{\mrm{BH}}$.  These merge in a turning point around  $T_\infty/T_{\mrm{BH}}\sim 0.93$, which suggests that Schwarzschild black droplets in equilibrium do not exist.

We use a novel numerical method to construct these geometries. It joins three existing numerical tools: transfinite interpolation on a Chebyshev grid, patching, and the DeTurck method. This method is not only useful for the construction of the solutions detailed here, but can be used in a broader sense with modest computational resources - see for instance \cite{Dias:2014cia} where this method was used to construct black rings in higher dimensions. In particular, the fact that we use transfinite interpolation on a Chebyshev grid means we do not require overlapping grids for the patching procedure\footnote{Overlapping grids are essential for patching using finite differences.}, which in turn not only simplifies the coding of the problem but also decreases the need for larger computational resources.

In the following section, we detail our numerical construction of these solutions.  In section 3, we investigate these solutions by computing embedding diagrams and the holographic stress tensor and matching our results to perturbation theory.  We make a few concluding remarks in section 4.  

\end{section}

%--- Constructing Black Droplets -------------------------
\begin{section}{Constructing Black Droplets over Planar Black Holes} 
%... Choosing a Reference Metric ...................
\begin{subsection}{Choosing a Reference Metric}
We opt to use the DeTurck method which was first introduced in \cite{Headrick:2009pv} and studied in great detail in \cite{Figueras:2011va}. This method alleviates issues of gauge fixing and guarantees the ellipticity of our equations of motion.  The method first requires a choice of reference metric $\bar g$ that is compatible with the boundary conditions.  One then solves the Einstein-DeTurck equation
\be\label{deturck}
R_{\mu\nu}=\frac{2\Lambda}{d-2}g_{\mu\nu}+\nabla_{(\mu}\xi_{\nu)}\;,
\ee
where $\xi^\mu=g^{\alpha\beta}\left(\Gamma^{\mu}_{\alpha\beta}+\bar\Gamma^{\mu}_{\alpha\beta}\right)$, and $\bar\Gamma^{\mu}_{\alpha\beta}$ is the Levi-Civita connection for $\bar g$.  For the kinds of solutions we are seeking, a maximal principle guarantees that any solution to \eqref{deturck} has DeTurck vector $\xi=0$, and is therefore also a solution to Einstein's equations \cite{Figueras:2011va}.

To find a black droplet suspended over  a planar black hole, the chosen reference metric must have a planar horizon, a droplet horizon, a symmetry axis, and a conformal boundary metric.  Furthermore, the reference metric must approach the planar black hole metric in the right limit.  Thus, the integration domain is schematically a pentagon.  Most numerical methods for PDEs use grids that lie on rectangular domains, but these methods can be extended to a pentagonal domain by patching two grids together.  Because of the difference in geometry between the two horizons, we will patch together two grids in different coordinate systems, each adapted to one of the horizons.

To motivate our choice of reference metric, let us first begin with $AdS_D$ in Poincar\'e coordinates
\be\label{poincare}
\mathrm{d}s^2_{AdS}=\frac{\ell^2}{z^2}\Big[-\mathrm{d}t^2+(\mathrm{d}z^2+\mathrm{d}r^2)+r^2 \mathrm{d}\Omega^2_{D-3}\Big]\;.
\ee
Notice that fixing the time and angular coordinates gives us a two-dimensional space that is confomally flat. This two-dimensional space in the line element \eqref{poincare} is written in Cartesian coordinates that can be adapted to a planar horizon.  We can also move to polar coordinates which are more suitable for a droplet horizon.  Therefore, we now search for a reference metric with a conformally flat subspace that also contains a droplet horizon and a planar horizon.  

To do this, let us first write the planar black hole in conformal coordinates.  We begin with the usual line element for the planar black hole solution in $D$ bulk dimensions:
\be\label{planar}
\mathrm{d}s^2_{\mrm{planar}}=\frac{\ell^2}{Z^2}\left[-\left(1-\frac{Z^{D-1}}{Z_0^{D-1}}\right)\mathrm{d}t^2+\frac{\mathrm{d}Z^2}{1-\frac{Z^{D-1}}{Z_0^{D-1}}}+\mathrm{d}r^2+r^2\mathrm{d}\Omega^2_{D-3}\right]\;.
\ee
Now let 
\be\label{toconformal}
\mathrm{d}z^2=\frac{\mathrm{d}Z^2}{1-\frac{Z^{d-1}}{Z_0^{d-1}}},
\ee
which gives us a line element of the form
\be\label{confplanar}
\mathrm{d}s^2_{\mrm{planar}}=\frac{\ell^2}{z^2 \tilde g(z)}\left[-\tilde f(z)(1-\tilde \lambda^2 z^2)^2\mathrm{d}t^2+\mathrm{d}z^2+\mathrm{d}r^2+r^2\mathrm{d}\Omega^2_{D-3}\right],
\ee
for some functions $\tilde g$, $\tilde f$, and constant $\tilde\lambda$. This line element has our desired conformal subspace.  For a boundary metric that is conformal to Schwarzschild, we find it numerically desirable to redefine the coordinates to 
\be\label{zrtoxy}
z^2=y\,,\qquad r=\frac{x}{1-x}\;,
\ee
which yields
\be\label{planarxy}
\mathrm{d}s^2_{\mrm{planar}}=\frac{\ell^2}{y\, g(y)}\left[-f_y(y)\;\mathrm{d}t^2+\frac{\mathrm{d}y^2}{4y}+\frac{\mathrm{d}x^2}{(1-x)^4}+\frac{x^2}{(1-x)^2}\mathrm{d}\Omega_2^2\right],
\ee
with
\be\label{fy}
 f_y(y)=f(y)(1-\lambda y)^2\;.
\ee
The planar horizon is located at the hyperslice $y=1/\lambda$.  The constant $\lambda$ (or $\tilde\lambda$) sets the temperature of the black hole and can be related to $Z_0$ in \eqref{planar}.  The functions $f$ and $g$ (or $\tilde f$ and $\tilde g$) are smooth, positive definite, and depend on the temperature.  They can be determined by integrating \eqref{toconformal} and inverting the resulting Hypergeometric function\footnote{Actually, we find it more convenient to determine $f$ and $g$ numerically by solving a set of ODEs rather than inverting the Hypergeometric.}.  To determine the integration constant, we choose $g(0)=f(0)=1$. 

Now let us write down a line element (not necessarily a solution of Einstein's equations) that has a single droplet horizon in conformal coordinates.  We search for something of the form
\be\label{confdroplet}
\mathrm{d}s^2_{\mrm{droplet}}=\frac{\ell^2}{z^2}\left[-\tilde f_\rho(\sqrt{z^2+r^2})\mathrm{d}t^2+\mathrm{d}z^2+\mathrm{d}r^2+r^2\mathrm{d}\Omega^2_{D-3}\right]\;,
\ee
where we have chosen $\tilde f_\rho$ to be a function of $\sqrt{z^2+r^2}$ in anticipation of moving to polar coordinates.  The function $\tilde f_\rho$ is determined by a choice of conformal boundary metric $\mathrm{d}s^2_\partial$.   At the boundary $z=0$, we must have
\be\label{feq}
-\tilde f_\rho(r)\mathrm{d}t^2+\mathrm{d}r^2+r^2 \mathrm{d}\Omega^2=\omega^2(r)\mathrm{d}s^2_\partial\;,
\ee
for some conformal factor $\omega$.  For a boundary metric that is conformal to Schwarzschild,
\be\label{schwarzschild}
\mathrm{d}s^2_\partial=-\left(1-\frac{R_0}{R}\right)\mathrm{d}\tau^2+\frac{\mathrm{d}R^2}{1-\frac{R_0}{R}}+R^2\mathrm{d}\Omega_2^2,
\ee
\eqref{feq} implies
\be
\tilde f_\rho(r)\mathrm{d}t^2=\frac{16\left(1-\frac{R_0}{r}\right)^2}{\left(1+\frac{R_0}{r}\right)^6}\mathrm{d}\tau^2\;.
\ee
We find that it is convenient to set $t=4\tau$.  This then uniquely specifies the function $\tilde f_\rho$, which together with \eqref{confdroplet} gives us our droplet line element in conformal coordinates.  Switching to the polar coordinates
\be\label{zrtorhoxi}
z=\frac{R_0}{\rho}\sqrt{1-\xi^2}\,,\qquad r=\frac{R_0\xi}{\rho}
\ee
gives us
\be\label{droplet0}
\mathrm{d}s^2_{\mrm{droplet}}=\frac{\ell^2}{1-\xi^2}\left[-\frac{\rho^2}{R_0^2}f_\rho(\rho)\mathrm{d}t^2+\frac{\mathrm{d}\rho^2}{\rho^2}+\frac{\mathrm{d}\xi^2}{1-\xi^2}+\xi^2\mathrm{d}\Omega_2^2\right]\;,
\ee
with
\be\label{frho}
f_\rho(\rho)=\frac{(1-\rho)^2}{(1+\rho)^6}\;.
\ee
By construction, the droplet horizon is at $\rho=1$ and its temperature (with respect to the time coordinate $\tau$) matches the temperature of the boundary Schwarzschild black hole.  Additionally, the line element \eqref{droplet0} can be used as a reference metric to reproduce the results of the solution in \cite{Figueras:2011va}.

Now we can attempt to combine the planar and droplet line elements to create our desired reference metric.  Guided by the similarities between \eqref{confplanar} and \eqref{confdroplet}, the reference metric we have chosen is
\begin{subequations}\label{referencemetric0}
\begin{align}
\mathrm{d}s_{\mrm{ref}}^2&=\frac{\ell^2}{y\;g}\left[-\frac{f_y f_\rho}{f_y+f_\rho-f_yf_\rho}\mathrm{d}t^2+\frac{\mathrm{d}y^2}{4y}+\frac{\mathrm{d}x^2}{(1-x)^4}+\frac{x^2}{(1-x)^2}\mathrm{d}\Omega_2^2\right]\\
&=\frac{\ell^2}{(1-\xi^2)\;g}\left[-\frac{\rho^2}{R_0^2}\frac{f_y f_\rho}{f_y+f_\rho-f_yf_\rho}\mathrm{d}t^2+\frac{\mathrm{d}\rho^2}{\rho^2}+\frac{\mathrm{d}\xi^2}{1-\xi^2}+\xi^2\mathrm{d}\Omega_2^2\right]\;,
\end{align}
\end{subequations}
where we treat $g$ and $f_y$ as functions of the coordinate $y$, and $f_\rho$ as a function of the coordinate $\rho$.  The $x$, $y$ coordinates are related to the $\rho$, $\xi$ coordinates through \eqref{zrtoxy} and \eqref{zrtorhoxi}:  
\begin{subequations}\label{eqs:coordref}
\begin{align}
x&=\frac{\xi}{\xi+\rho/R_0}\,, \qquad\qquad\;\; y=\frac{R_0^2}{\rho^2}(1-\xi^2)\,,\\
\frac{\rho^2}{R_0^2}&=\frac{(1-x)^2}{x^2+(1-x)^2y}\,,\qquad \xi^2=\frac{x^2}{x^2+(1-x)^2y}\;.
\end{align}
\end{subequations}  

The reference metric \eqref{referencemetric0} has a regular planar horizon at $y=1/\lambda$, a regular droplet horizon at $\rho=1$, and an axis at $x=0$ (or $\xi=0$).  Near $x=1$, we recover the planar black hole metric as written in \eqref{planarxy}.  Since $g(0)=f(0)=1$, near $y=0$ or $\xi=1$ we have (in the $\rho$, $\xi$ coordinate system)
\be
\mathrm{d}s^2_{\mrm{ref}}\rightarrow\frac{\ell^2}{1-\xi^2}\left[\frac{\mathrm{d}\xi^2}{1-\xi^2}+\frac{R_0^2(1+\rho)^4}{16\rho^2}\mathrm{d}s_\partial^2\right]\;,
\ee
where
\be\label{boundaryrho}
\mathrm{d}s_\partial^2=-\frac{(1-\rho)^2}{16(1+\rho)^2}\mathrm{d}t^2+\frac{R_0^2(1+\rho)^4\mathrm{d}\rho^2}{16\rho^4}+\frac{R_0^2}{16\rho^2}(1+\rho)^4\mathrm{d}\Omega_2^2\;.
\ee
We can see that this is equivalent to Schwarzschild \eqref{schwarzschild} by performing the coordinate transformation
\be
t=4\tau\,,\qquad r=-1+2\frac{R}{R_0}\left(1-\sqrt{1-\frac{R_0}{R}}\right)\;.
\ee

We have thus found a reference metric that is compatible with our desired boundary conditions.  By construction, this reference metric can be written in two orthogonal coordinate systems, with all boundaries in our domain being a constant hyperslice in at least one of these two coordinate systems.  Furthermore, in the $\lambda\rightarrow0$ limit, our reference metric becomes the droplet metric \eqref{droplet0}, which is an appropriate reference metric for a droplet without a planar black hole.  

We have two parameters given by $\lambda$ and $R_0$, which determine the temperatures $T_\infty$ and $T_\mrm{BH}$, respectively.  This system, however, only has one dimensionless parameter given by the ratio $T_\infty/ T_\mrm{BH}$, so we have one remaining gauge degree of freedom which we can choose for numerical convenience.  

\end{subsection}
%... Ansatz and Boundary Conditions ...................
\begin{subsection}{Ansatz and Boundary Conditions}
With a reference metric in hand, we can now write down a metric ansatz:
\begin{subequations}\label{referencemetric}
\begin{align}
\mathrm{d}s^2&=\frac{\ell^2}{y\;g}\Bigg\{-\frac{f_y f_\rho}{f_y+f_\rho-f_yf_\rho}T_c\;\mathrm{d}t^2+\frac{A_c\;\mathrm{d}y^2}{4y}\nonumber\\
&\qquad\qquad\qquad+\frac{B_c}{(1-x)^4}\left[\mathrm{d}x+\frac{x(1-x)^3F_c}{x^2+(1-x)^2y}\mathrm{d}y\right]^2+\frac{x^2S_c}{(1-x)^2}\mathrm{d}\Omega_2^2\Bigg\}\label{subeq:xy}\\
&=\frac{\ell^2}{(1-\xi^2)\;g}\Bigg\{-\frac{\rho^2}{R_0^2}\frac{f_y f_\rho}{f_y+f_\rho-f_yf_\rho}T_p\;\mathrm{d}t^2+\frac{A_p\;\mathrm{d}\rho^2}{\rho^2}\nonumber\\
&\qquad\qquad\qquad\qquad\qquad+\frac{B_p}{1-\xi^2}\left[\mathrm{d}\xi+\frac{\xi}{\rho}(1-\xi^2)F_p\;\mathrm{d}\rho\right]^2+\xi^2\mathrm{d}\Omega_2^2\Bigg\}\label{subeq:rhoxi}\;,
\end{align}
\end{subequations}
where $T_c$, $A_c$, $B_c$, $F_c$, and $S_c$ are functions of the Cartesian coordinates $x$ and $y$, and $T_p$, $A_p$, $B_p$, $F_p$, and $S_p$ are functions of the polar coordinates $\rho$ and $\xi$.  Since we must demand that the metric is equivalent between these two coordinate systems, the functions are related to each other via
\begin{align}\label{matching}
T_c=T_p\,,\qquad S_c=S_p\,,\qquad A_c=\frac{A_pB_p}{A_p\xi^2+B_p(1-\xi^2)(1-F_p\xi^2)^2}\,,\nonumber\\
B_c=A_p\xi^2+B_p(1-\xi^2)(1-F_p\xi^2)^2\,,\qquad F_c=\frac{A_p-B_p(1-F_p\xi^2)(1+F_p(1-\xi^2))}{2(A_p\xi^2+B_p(1-\xi^2)(1-F_p\xi^2)^2)}\,,
\end{align}
where we used the coordinate transformations (\ref{eqs:coordref}).

Now let us discuss boundary conditions.  At the boundary $y=0$ or $\xi=1$, we must recover a metric conformal to Schwarzchild.  This was already done in the reference metric, so we choose
\begin{subequations}
\begin{align}
T_c|_{y=0}=A_c|_{y=0}=B_c|_{y=0}=S_c|_{y=0}=1\,,\qquad F_c|_{y=0}=0\;,\\
T_p|_{\xi=0}=A_p|_{\xi=0}=B_p|_{\xi=0}=S_p|_{\xi=0}=1\,,\qquad F_p|_{\xi=0}=0\;.
\end{align}
\end{subequations}
Similarly, we must recover the planar black hole at $x=1$ and impose 
\be
T_c|_{x=1}=A_c|_{x=1}=B_c|_{x=1}=S_c|_{x=1}=1\,,\qquad F_c|_{x=1}=0\;.
\ee
The remaining boundary conditions are determined by regularity.  At the planar horizon $y=1/\lambda$, we need
\begin{align}
&T_c|_{y=1/\lambda}=A_c|_{y=1/\lambda}\,,\qquad F_c|_{y=1/\lambda}=0\,,\nonumber\\
\partial_yT_c|_{y=1/\lambda}&=\partial_yA_c|_{y=1/\lambda}=\partial_yB_c|_{y=1/\lambda}=\partial_yS_c|_{y=1/\lambda}=0\;.
\end{align}
At the axis, $x=0$ or $\xi=0$, we require
\begin{subequations}
\begin{align}
&B_c|_{x=0}=S_c|_{x=0}\;,\qquad \partial_xT_c|_{x=0}=\partial_xA_c|_{x=0}=\partial_xB_c|_{x=0}=\partial_xS_c|_{x=0}=\partial_xF_c|_{x=0}=0\;,\\
&B_p|_{\xi=0}=S_p|_{\xi=0}\;,\qquad \partial_\xi T_p|_{\xi=0}=\partial_\xi A_p|_{\xi=0}=\partial_\xi B_p|_{\xi=0}=\partial_\xi S_p|_{\xi=0}=\partial_\xi F_p|_{\xi=0}=0\;.
\end{align}
\end{subequations}
Finally, at the droplet horizon $\rho=1$, 
\begin{align}
&T_p|_{\rho=1}=A_p|_{\rho=1}\,,\qquad F_p|_{\rho=1}=0\,,\nonumber\\
\partial_\rho T_p|_{\rho=1}&=-\frac{2R_0^2(1-\xi^2)g'A_p(3B_pS_p+A_p(2B_p+S_p))}{3gB_pS_p}\bigg|_{\rho=1}\,,\nonumber\\
\partial_\rho A_p|_{\rho=1}&=-\frac{2R_0^2(1-\xi^2)g'A_p(3B_pS_p+2A_p(2B_p+S_p))}{3gB_pS_p}\bigg|_{\rho=1}\,,\nonumber\\
\partial_\rho B_p|_{\rho=1}&=-\frac{2R_0^2(1-\xi^2)g'}{g}B_p\bigg|_{\rho=1}\,,\nonumber\\
\partial_\rho S_p|_{\rho=1}&=-\frac{2R_0^2(1-\xi^2)g'}{g}S_p\bigg|_{\rho=1}\;.
\end{align}
\end{subsection}

%... Numerics ...................
\begin{subsection}{Numerics}

\begin{figure}[th]
\begin{center}
\includegraphics[width=.42\textwidth]{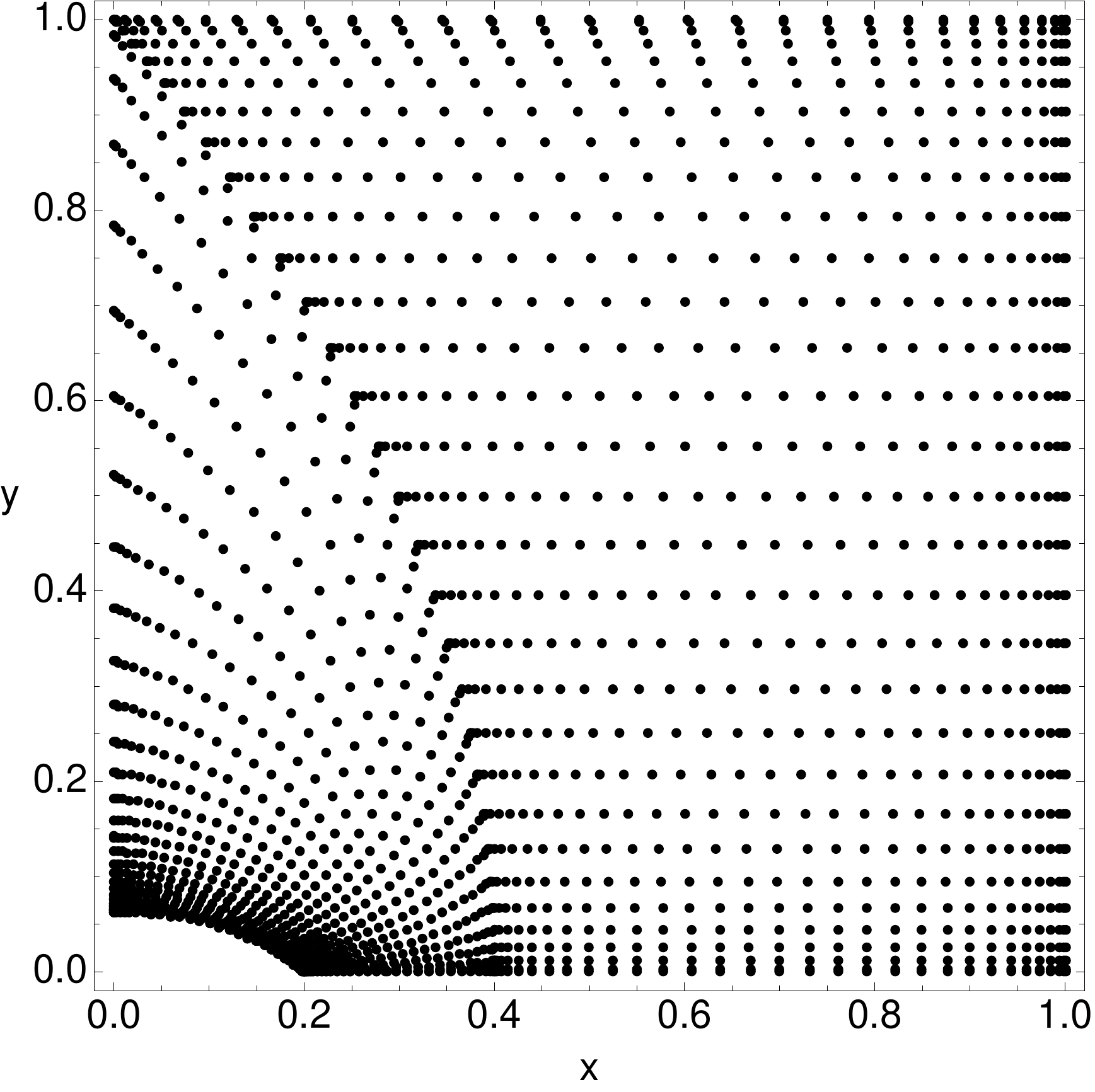}
\end{center}
\caption{A grid for our computational domain formed by combining transfinite interpolation and patching.  We work with one patch in `polar' coordinates and the other in the `cartesian' coordinates shown here.  }\label{Fig:grid}
\end{figure}  

To solve the equations of motion numerically, we employ a standard Newton-Raphson relaxation algorithm using pseudospectral collocation.  To choose a suitable grid, we first divide the entire integration domain into two patches, one in each coordinate system.  We then place a spectral grid on each patch using transfinite interpolation on a Chebyshev grid.  An example of such a grid is shown in figure Fig.~\ref{Fig:grid}.  In addition to imposing the boundary conditions, we require the smoothness of the metric across patches.  This amounts to requiring \eqref{matching} and the equivalent expression for normal derivatives across the patch boundary.  We obtained our first solution by using the reference metric as a Newton-Raphson seed.  

\begin{figure}[th]
\begin{center}
\includegraphics[width=.42\textwidth]{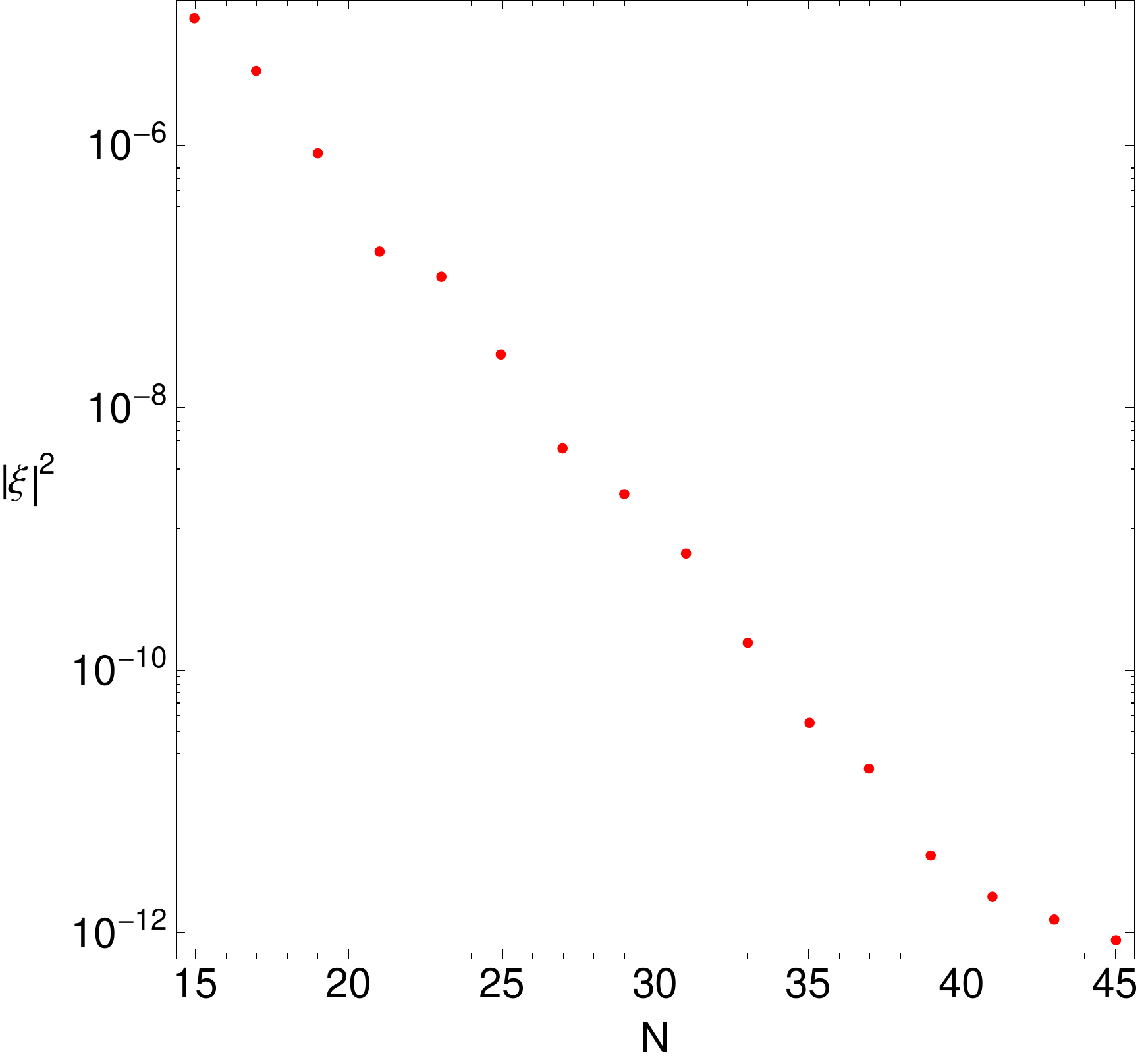}
\end{center}
\caption{The error in the deTurck norm as a function of the grid size $(N+N)\times N$ for one of our droplet solutions.  We see an exponential convergence down to machine error of $\sim 10^{-11}$.}\label{Fig:convergence}
\end{figure}  

Since it has been proven that the DeTurck vector $\xi=0$ for any solution of \eqref{deturck} satisfying boundary conditions such as those appearing here \cite{Figueras:2011va}, we can use this quantity to monitor our numerical error and test the convergence of our code.  As seen in Fig.~\ref{Fig:convergence}, our numerics converge exponentially with increasing grid size, as predicted by pseudospectral methods.   All of our results presented below have $|\xi|^2<10^{-10}$.  We have also verified that our results do not change when we vary the location of our patch boundary or when we change $\lambda$ and $R_0$ while keeping $T_\infty/ T_\mrm{BH}$ fixed.
\end{subsection}
\end{section}

%--- Results -------------------------
\begin{section}{Results}
%... Embedding and Distance Between the Horizons ...................
\begin{subsection}{Embedding and Distance Between the Horizons}

\begin{figure}[th]
\begin{center}
\includegraphics[width=.42\textwidth]{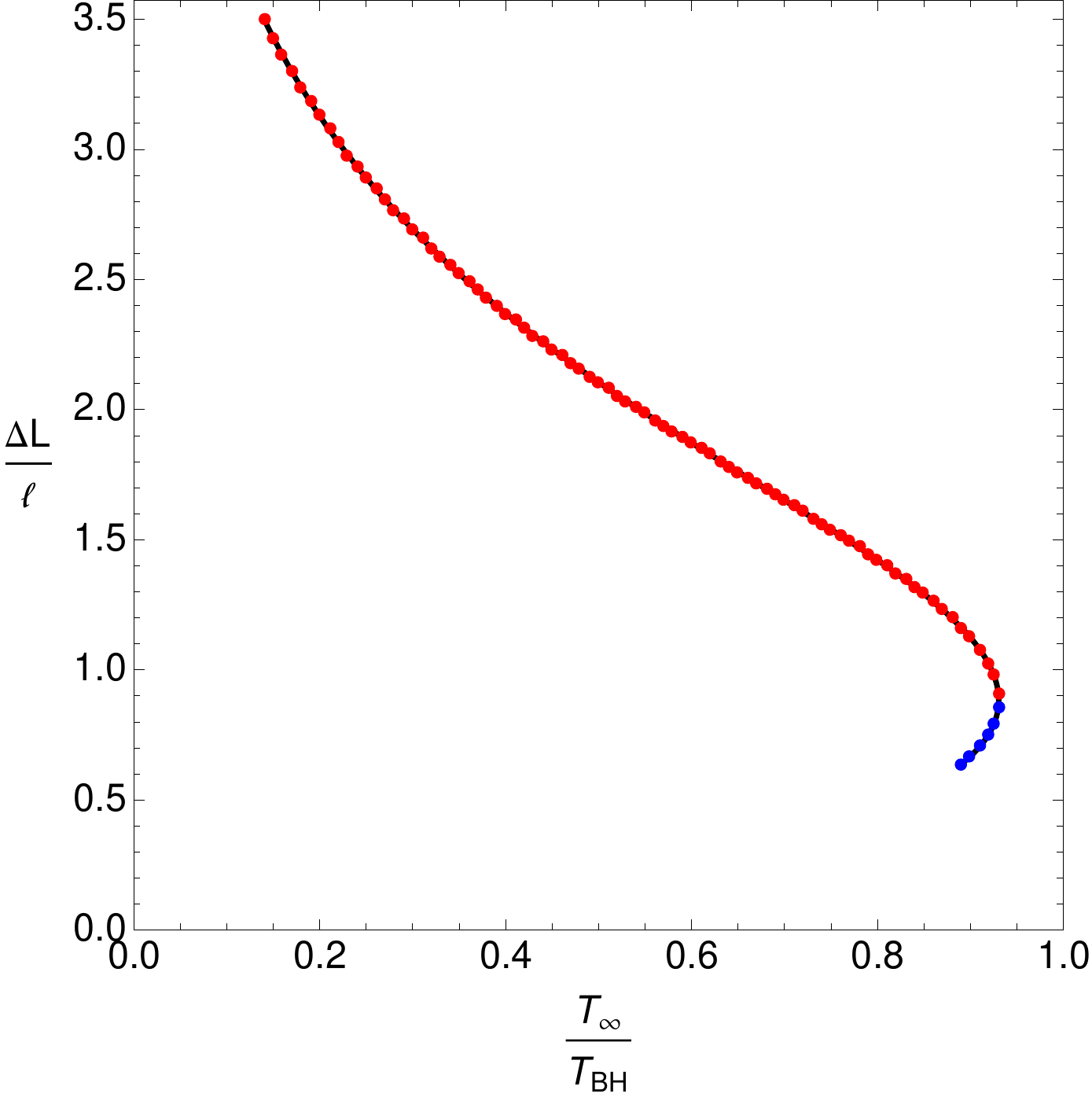}
\end{center}
\caption{The proper length between the droplet and planar horizons along the axis of symmetry as a function of the temperature ratio.  For a given temperature ratio, there can be two droplet solutions.  The turning point occurs around $T_\infty/ T_\mrm{BH}\sim0.93$, which suggests that the equilibrium solution does not exist.}\label{Fig:length}
\end{figure}  

To get a sense for the relationship between these two horizons, in Fig.~\ref{Fig:length} we plot the proper distance between the horizons along the axis of symmetry as a function of temperature.  For small $T_\infty/ T_\mrm{BH}$, there are solutions with a large distance between the black droplet and the planar black hole.  These are solutions which are close to the $T_\infty=0$ solution found in \cite{Figueras:2011va}.  As we follow these solutions with \emph{increasing}  $T_\infty/ T_\mrm{BH}$, we find that the proper distance decreases until $T_\infty/ T_\mrm{BH}\sim0.93$.  At this value there is a turning point where the proper distance continues to decrease only if we  \emph{decrease}  $T_\infty/ T_\mrm{BH}$.  These results suggest that $T_\infty/ T_\mrm{BH}\sim0.93$ is a critical temperature above which only (possibly flowing) funnel solutions exist.  In particular, the equilibrium state would be the funnel solution found in \cite{Santos:2012he}.

To help us understand the geometry of the solutions, we embed the two horizons in Euclidean hyperbolic space:
\be\label{hyperbolicspace}
\mathrm{d}s_{\mbb H}^2=\frac{\ell^2}{z^2}\left(\mathrm{d}z^2+\mathrm{d}r^2+r^2 \mathrm{d}\Omega^2_{D-3}\right)\;.
\ee
Demanding that the pullback of hyperbolic space to a curve $\gamma(x)=(z(x),r(x))$ is equal to the pullback of our solution to the horizon gives a system of ODEs in $z(x)$ and $r(x)$.  We solve these ODEs numerically to obtain our embedding diagram.  

\begin{figure}[th]
\begin{center}
\includegraphics[width=.42\textwidth]{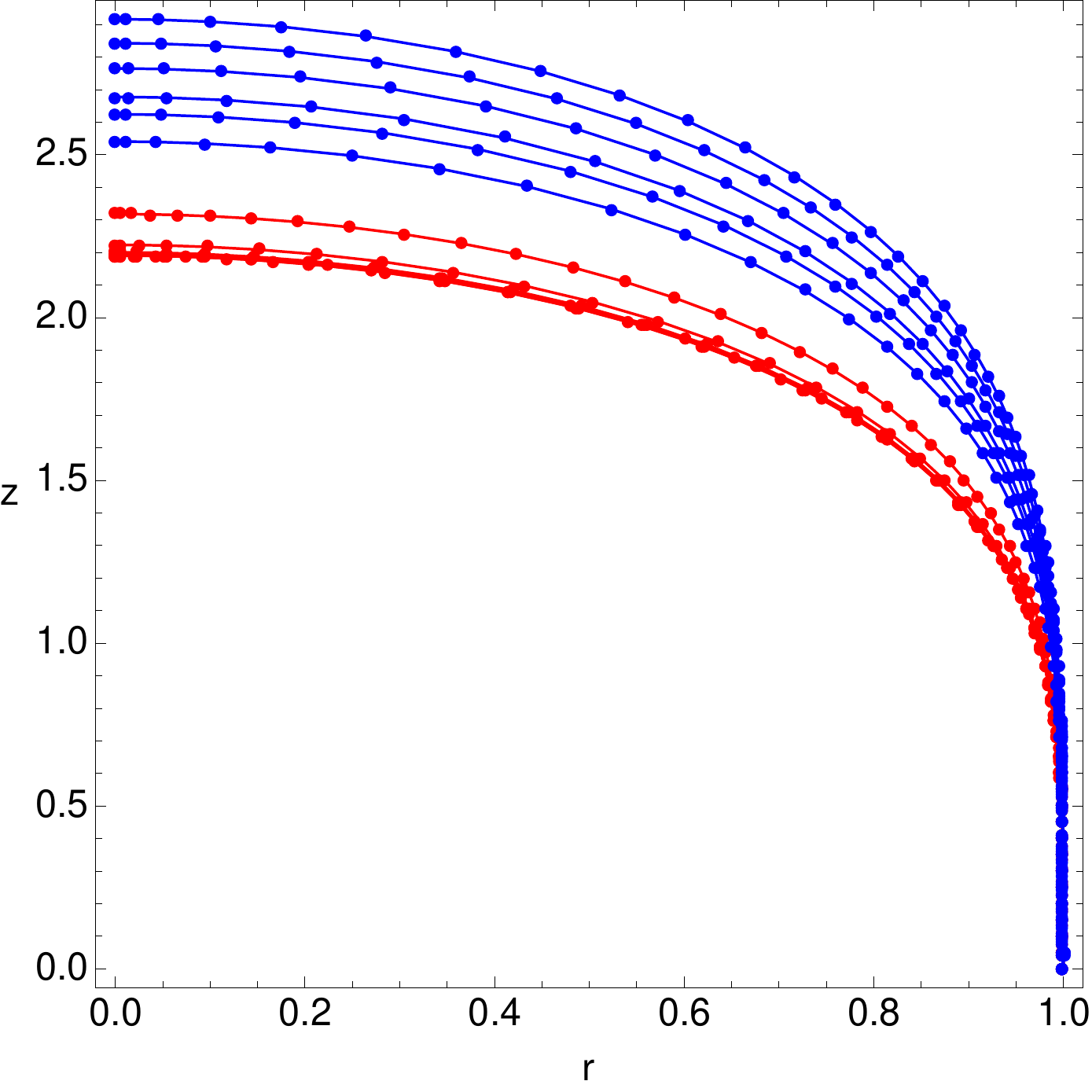}\qquad\qquad
\includegraphics[width=.42\textwidth]{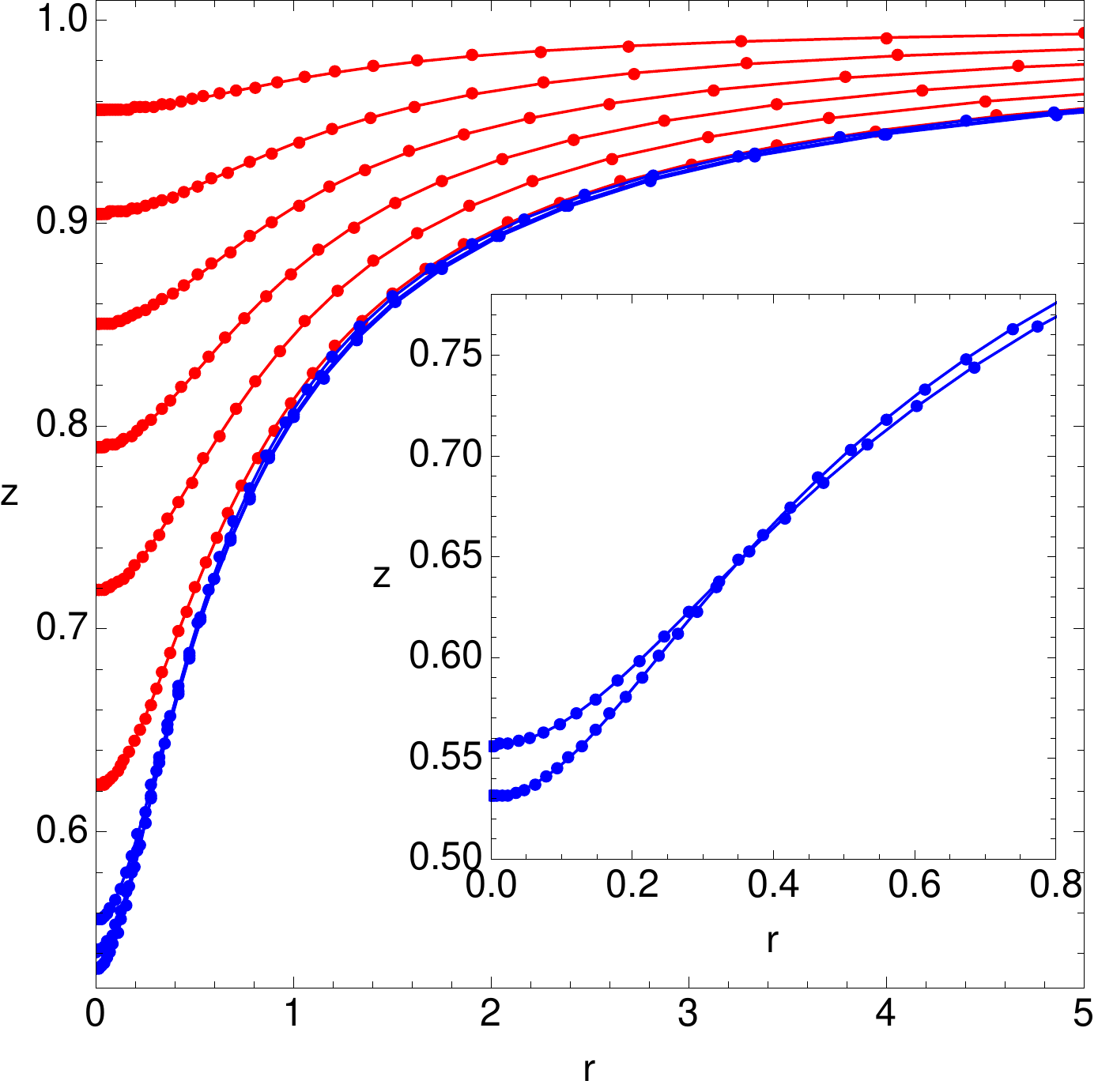}
\end{center}
\caption{The embeddings of droplet horizons (left), and planar horizons (right) in hyperbolic space \eqref{hyperbolicspace}.  The droplet horizons are normalised to $r=1$ at the boundary, and the planar horizons are normalised to $z=1$ at $r\rightarrow\infty$.  The blue curves are long droplet solutions and the red curves are for short droplet solutions.  The inset plot on the right is a zoomed in plot for two of the long-droplet solutions.}\label{Fig:embedding}
\end{figure}  

The embeddings of the droplet horizon and planar horizon are shown in Fig.~\ref{Fig:embedding}.  The size of the droplets at the boundary is normalised to 1, and the location of the planar black hole far from the droplet is also normalised to 1.  Starting at small $T_\infty/ T_\mrm{BH}$, the droplet horizon looks very similar to that of \cite{Figueras:2011va}, and the planar horizon is approximately flat.  As we increase $T_\infty/ T_\mrm{BH}$, we see that even past the turning point, the droplet horizon continues to lower itself deeper into the bulk and the centre of the planar horizon continues to rise towards the boundary.  Based on the shape of these solutions from the embedding diagram, we call our two branches of droplet solutions \emph{long dropets} and \emph{short droplets}.  Similar behaviour has been observed for black droplets in global AdS \cite{donsantos}.

Eventually, our numerics break down and we are unable to continue the long droplets any further.  We can only conjecture a number of possibilities.  One scenario is that the long droplets continue to exist down to $T_\infty=0$,  these solutions may join with the $AdS$ black string.  In this case, one might reinterpret the naked singularity of the string as a degenerate droplet/funnel merger point.  

Another possibility is that the two horizons merge at some finite temperature ratio towards a funnel.  At the merger, they would reach a conical transition. Since the two horizons are not at the same temperature, this would mean a transition between a static solution to a stationary one with some amount of flow.  But going a small amount across a conical merger should not change the geometry far from the cone significantly, so the amount of heat flux at infinity should be small.  If this picture is correct, this would mean that there are two types of flowing funnel solutions, one with a narrow neck and small flow, and one with a wider neck with larger flow.  Though, like the caged black holes \cite{Kudoh:2004hs}, it is also possible that there is no stationary solution on the funnel side of the merger, and the solution necessarily becomes dynamical and possibly evolves into a wide flowing funnel.  

\bigskip

\end{subsection}
%... Stress Tensor...................
\begin{subsection}{Stress Tensor}
Now we compute the boundary stress tensor.  The procedure we use is similar to those of \cite{deHaro:2000xn}.  We expand the equations of motion off of the boundary in a Fefferman-Graham expansion, choosing a conformal frame that gives Schwarzschild on the boundary.  We can then read off the stress tensor from one of the higher order terms in the expansion.  There is no conformal anomaly in our case because we have chosen a boundary metric that is Ricci flat.  

\begin{figure}[ht]
\begin{center}
\includegraphics[width=.42\textwidth]{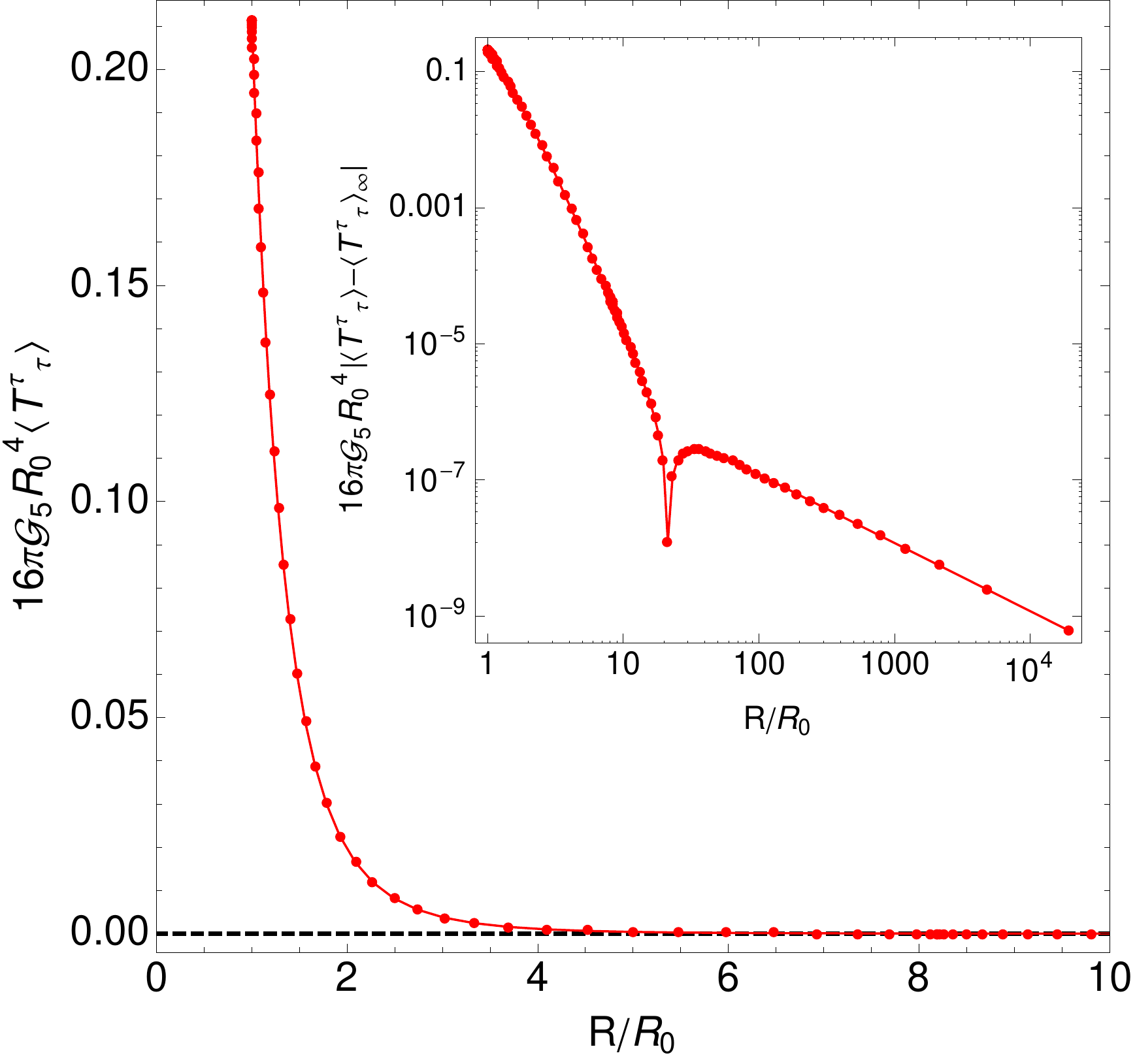}\qquad
\includegraphics[width=.42\textwidth]{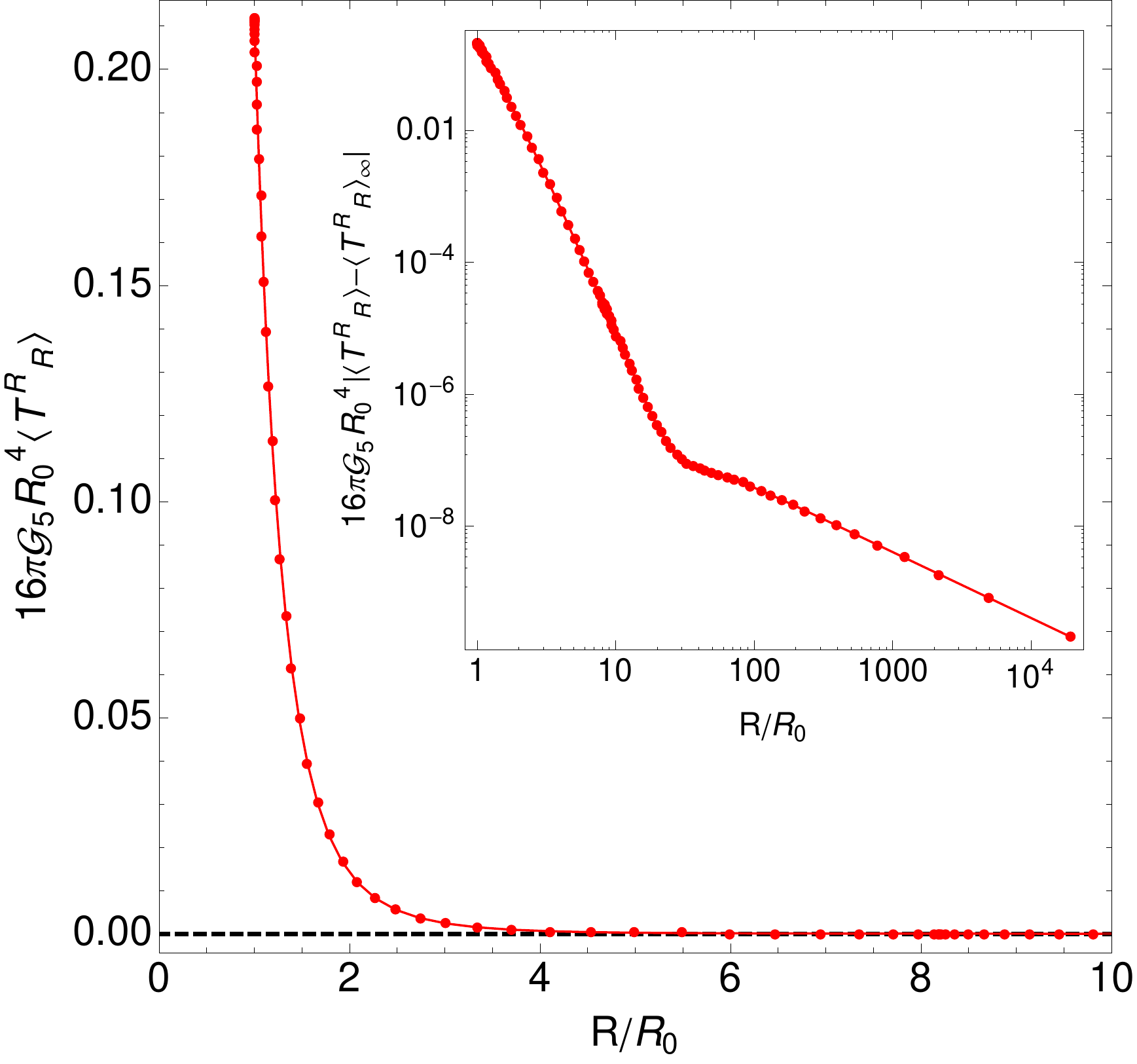}
\end{center}
\caption{Components for the stress tensor with $T_{\infty}/T_\mrm{BH}= 0.15$.  The dashed black line is the value of the stress tensor for the planar black hole.  The insets are log-log plots with this asymptotic value subtracted (the kinks appear because of the absolute value).  }\label{Fig:stress15}
\end{figure}  

\begin{figure}[ht]
\begin{center}
\includegraphics[width=.42\textwidth]{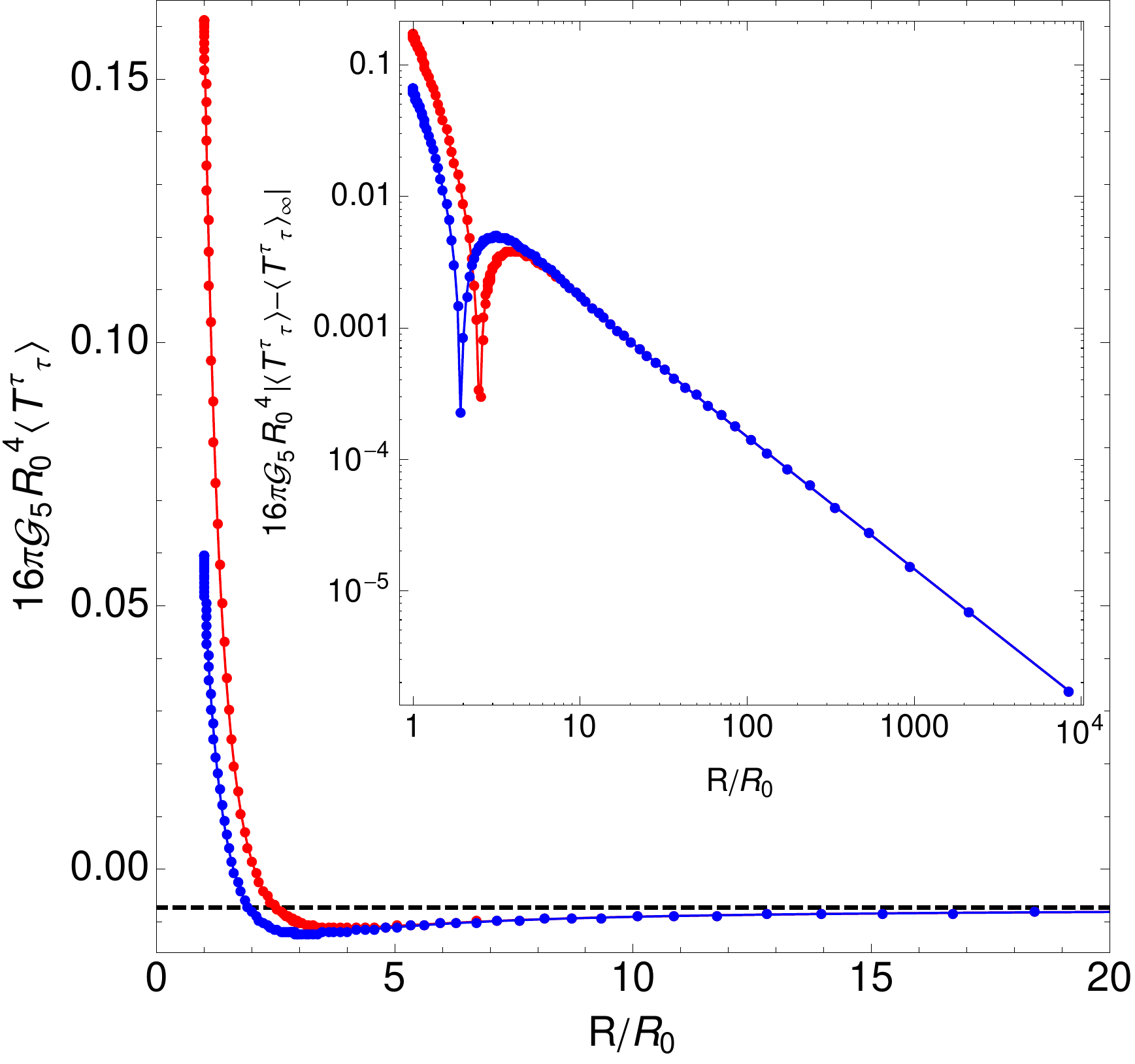}\qquad
\includegraphics[width=.42\textwidth]{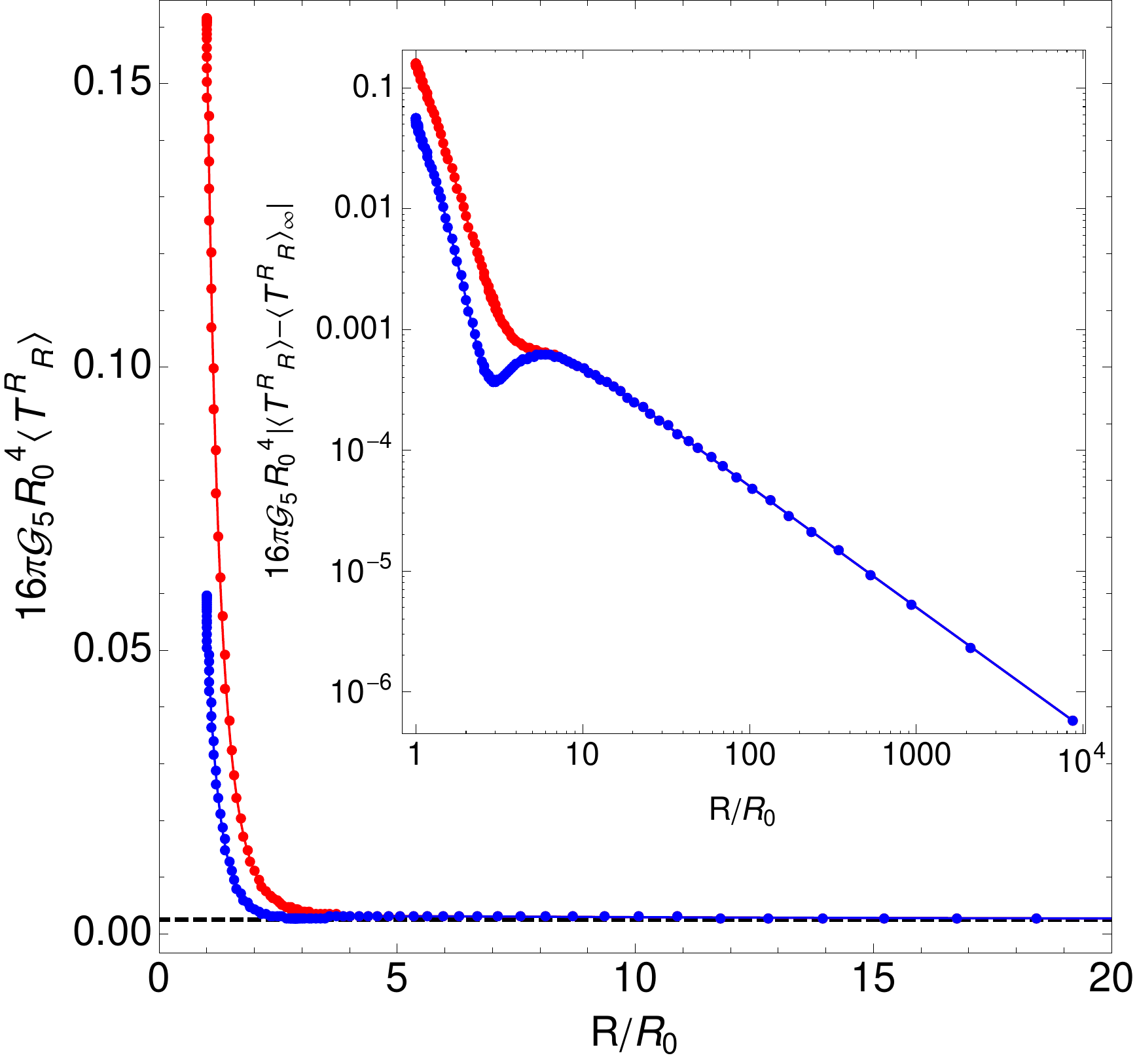}

\end{center}
\caption{Components for the stress tensor with $T_{\infty}/T_\mrm{BH}= 0.89$.  (Same scheme as Fig.~\ref{Fig:stress15}.)  The larger red curve is the short droplet while the smaller blue curve is the long droplet.}\label{Fig:stress89}
\end{figure}  

Representative stress tensors of our solutions are plotted in Figs.~\ref{Fig:stress15}, and \ref{Fig:stress89}.  Far from the boundary black hole, the stress tensor fits the form
\be
\langle T_{\mu\nu}\rangle\sim k_0+\frac{k_1}{R}+O\left(\frac{1}{R^2}\right)\;,
\ee
where $k_0$ is the boundary stress tensor for a bulk planar black hole.  This $R^{-1}$ behaviour was also found for the funnel solutions in \cite{Santos:2012he}.  

In the insets of Figs.~\ref{Fig:stress15}, and \ref{Fig:stress89}, we subtract $k_0$ from the stress tensor, take an absolute value, and plot the result using a Log-Log scale.  Note that there are clearly two power-law regimes.  Far from the black hole, we see a $R^{-1}$ power law, similar to that of a funnel.  Closer to the black hole, we see a $R^{-5}$ power law, similar to that of the droplets found in \cite{Figueras:2011va}.  

This dual power-law can be explained from the bulk perspective.  The presence of the droplet warps the planar horizon, making it funnel-like far away.  This is most easily seen in our embedding diagrams in Fig. \ref{Fig:embedding}.  This funnel-like behaviour gives the stress tensor a $R^{-1}$ power law.  Closer to the droplet, the physics near the boundary is dominated by the hotter droplet horizon rather than the planar horizon, giving a $R^{-5}$ droplet behaviour.  As the distance between the horizons decreases, this $R^{-5}$ behaviour becomes more obscured.  

In Fig.~\ref{Fig:stress89} we can see that both long and short droplets have the same large $R$ behaviour, suggesting that this is universal.  Indeed, we shall match this behaviour with perturbation theory in the next section.

\end{subsection}

%... Perturbation Theory...................

\begin{subsection}{Matching with perturbation theory}
Far away from the axis of symmetry of the droplet, \emph{i.e.} close to $x=1$ in Eq.~(\ref{subeq:xy}), perturbation theory should be valid. This region can solely be studied using standard perturbation theory techniques around the planar black hole line element (\ref{planar}). For concreteness, we will take $D=5$, even though our procedure admits a straightforward extension to arbitrary $D$.

We first note that the planar black hole can be written as
\begin{equation}
\mathrm{d}s^2_{\mrm{planar}}=\frac{\ell^2}{Z^2}\left[-\left(1-\frac{Z^4}{Z_0^4}\right)\mathrm{d}t^2+\frac{\mathrm{d}Z^2}{1-\frac{Z^4}{Z_0^4}}+\mathrm{d}\mathbb{E}_{3}^2\right]\,,
\end{equation}
where $\mathrm{d}\mathbb{E}_{3}^2$ is the line element of three dimensional Euclidean space. Following \cite{Kodama:2003jz}, we can decompose our perturbations according to how they transform under diffeomorphisms of $\mathbb{E}_3$. These can be decomposed as tensors, vectors or scalar derived perturbations. Here, we are primarily interested in scalar perturbations. Its basic building block are the scalar harmonics on $\mathbb{E}_3$, which satisfy the following simple equation
\begin{equation}
\Box_{\mathbb{E}_3} \mathbb{S}+\alpha^2 \mathbb{S}=0\,.\nonumber
\end{equation}
Furthermore, we are interested in perturbations that do not break the $2-$sphere inside $\mathbb{E}_3$, so we only have radial dependence in $\mathbb{S}$. These can be computed and we find
\begin{equation}
\mathbb{S}(r) = C_1 \frac{\sin (\alpha\,R)}{R}+C_2 \frac{\cos (\alpha\,R)}{R}\,.\nonumber
\end{equation}
A general perturbation can be decomposed as
\begin{equation}
h_{ab} = f_{ab}(t,Z)\mathbb{S},\qquad h_{aI} = f_a(t,Z) \nabla_I \mathbb{S}\qquad h_{IJ} = H_L(t,Z) g_{IJ}+H_T(t,Z) \left(\nabla_I \nabla_J \mathbb{S}+\alpha^2\frac{g_{IJ}}{3}\mathbb{S}\right)\,,
\end{equation}
where lower case latin indices run over $\{t,Z\}$ and upper case latin indices run over coordinates in $\mathbb{E}_3$. In addition, we are interested in non-normalizable perturbations that are time independent. This means we can set $f_{tZ}=f_t=0$. We are thus left with two gauge degrees of freedom, corresponding to reparametrizations of $Z$ and $R$. We fix this by demanding $f_Z=0$ and $H_T=0$. We are thus left with three variables: $f_{tt}(Z)$, $f_{ZZ}(Z)$ and $H_L(Z)$. The Einstein equations automatically fix $f_{ZZ}$ as an algebraic function of $f_{tt}$ and $H_L$:
\begin{equation}
f_{ZZ} = \frac{f_{tt}}{\left(1-\frac{Z^4}{Z_0^4}\right)^2}-\frac{H_L}{\left(1-\frac{Z^4}{Z_0^4}\right)}\,.\nonumber
\end{equation}
The remaining Einstein equations reduce to two first order equations in $H_L$ and $f_{tt}$, which we reduce to a single second order equation in $f_{tt}$:
\begin{multline}
f_{tt}^{\prime\prime}(w)+\frac{4 \alpha ^2 w_0^2 w^3-6 w^4+w_0^4 \left(6+4 \alpha ^2 w\right)}{w\left[3 w^4+2 w_0^2 w^2 \left(3-2 \alpha ^2 w\right)+w_0^4 \left(3+4 \alpha ^2 w\right)\right]}f_{tt}^{\prime}(w)+\\
 \frac{24 w^5-w^3 w_0^2 \left(96+\alpha ^2 w\right)+\alpha ^2 w_0^6 \left(15+4 \alpha ^2 w\right)-2 w w_0^4 \left[\alpha ^2 w \left(9+2 \alpha ^2 w\right)-36\right]}{4 w \left(w^2-w_0^2\right) \left[3 w^4+2 w_0^2 w^2 \left(3-2 \alpha ^2 w\right)+w_0^4 \left(3+4 \alpha ^2 w\right)\right]}f_{tt}(w)=0\,,
 \label{eq:comp}
\end{multline}
where we performed the coordinate transformation $Z^2=w$ and defined $Z_0^2=w_0$. Before proceeding to determine the solution, let us first discuss the boundary conditions. Recall that at the boundary we need to recover the Schwarzschild line element (\ref{schwarzschild}) expanded at large values of $R$. This is equivalent to demanding:
\begin{equation}
\lim_{Z\to0}h_{tt}(Z,R) = \frac{\ell^2}{Z^2}\frac{R_0}{R}\,.
\label{eq:bc1}
\end{equation}
This boundary condition picks $\alpha=0$, and without loss of generality we take $C_2=\ell^2$. For this choice, Eq.~(\ref{eq:comp}) admits a simple analytic solution:
\begin{equation}
f_{tt}(Z) = B  \left(Z^4+Z_0^4\right)+A  \left(\frac{Z_0^4}{Z^2}+Z^2\right)\,,
\label{eq:ftt}
\end{equation}
where $A$ and $B$ are constants to be chosen in what follows. Regularity at the black hole horizon and the boundary condition (\ref{eq:bc1}) demand $A = R_0/Z_0^4$ and $B = -R_0/Z_0^6$.

The full metric perturbation can be reconstructed from Eq.~(\ref{eq:ftt}) and is given by:
\begin{equation}
h_{\mu\nu} = 
\frac{\ell^2}{Z^2}\left[
\begin{array}{ccccc}
 -\frac{R_0 \left(Z^2-Z_0^2\right) \left(Z^4+Z_0^4\right)}{R Z_0^6} & 0 & 0 & 0 & 0 \\
 0 & \frac{2 Z^2 R_0 Z_0^4}{R \left(Z^2-Z_0^2\right) \left(Z^2+Z_0^2\right)^2} & 0 & 0 & 0 \\
 0 & 0 & \frac{R_0 \left(Z^2+Z_0^2\right)}{R Z_0^2} & 0 & 0 \\
 0 & 0 & 0 & \frac{R R_0 \left(Z^2+Z_0^2\right)}{Z_0^2} & 0 \\
 0 & 0 & 0 & 0 & \frac{R R_0 \left(Z^2+Z_0^2\right)}{Z_0^2} \sin^2\theta\\
\end{array}
\right]\label{eq:hmetric}\,,
\end{equation}
where we parametrize the $2-$sphere in the standard way $\mathrm{d}\Omega^2_2 = \mathrm{d}\theta^2+\sin^2\theta\mathrm{d}\phi^2$. This metric perturbation does not seem to have a boundary metric perturbation that approaches the large $R$ behavior of the Schwarzschild line element (\ref{schwarzschild}). However, this is an illusion of the gauge we choose to work in. If we perform a gauge transformation with gauge parameter $\xi = -\ell^2 R_0/(2\,Z^2)\,dR$, we bring the metric perturbation (\ref{eq:hmetric}) to
\begin{multline}
h^{F}_{\mu\nu} \equiv h_{\mu\nu}+2\nabla_{(\mu}\xi_{\nu)}= \\
\frac{\ell^2}{Z^2}\left[
\begin{array}{ccccc}
 -\frac{R_0 \left(Z^2-Z_0^2\right) \left(Z^4+Z_0^4\right)}{R Z_0^6} & 0 & 0 & 0 & 0 \\
 0 & \frac{2 Z^2 R_0 Z_0^4}{R \left(Z^2-Z_0^2\right) \left(Z^2+Z_0^2\right){}^2} & 0 & 0 & 0 \\
 0 & 0 & \frac{Z^2 R_0 \left(\frac{1}{Z_0^2}+\frac{1}{Z^2}\right)}{R} & 0 & 0 \\
 0 & 0 & 0 & \frac{R Z^2 R_0}{Z_0^2} & 0 \\
 0 & 0 & 0 & 0 & \frac{R Z^2 R_0}{Z_0^2}\sin^2\theta \\
\end{array}
\right]\label{eq:hmetricf}\,,
\end{multline}
which manifestly exhibits the boundary metric we desire.

It is now a simple exercise to determine the perturbed stress energy tensor in terms of the boundary black hole temperature $T_\mrm{BH}$ and planar temperature $T_{\infty}$:

\begin{subequations}
\begin{align}
16 \pi G{R_0}^4\langle \delta T^{t}_{\phantom{t}t}\rangle = -\frac{3}{256}\left(\frac{T_{\infty}}{T_\mrm{BH}}\right)^4 \left(1+\frac{2R_0}{R}+\ldots\right)\,,\\
\langle \delta T^{R}_{\phantom{R}R}\rangle=\langle \delta T^{\theta}_{\phantom{\theta}\theta}\rangle=\langle \delta T^{\phi}_{\phantom{\phi}\phi}\rangle=-\frac{1}{3}\langle \delta T^{t}_{\phantom{t}t}\rangle \,.
\end{align}
\label{eq:perturb}
\end{subequations}

This should be the leading asymptotic behavior of the holographic stress energy tensor of the droplet solution as we approach $R\to+\infty$. This is partially confirmed by \cite{Santos:2012he} where the stress energy tensor is found to be consistent with (\ref{eq:perturb}) if $T_{\infty} = T_\mrm{BH}=T_{\mathrm{Schwarzschild}}$.  A linear fit of our log-log plots agrees with (\ref{eq:perturb}) to less than $0.1\%$.  

The next correction should appear at $\mathcal{O}(R^{-2})$ and can be computed using a similar approach, albeit with a more tedious calculation. Based on our solution at smaller $R$, we expect the first undetermined coefficient in the $R=+\infty$ expansion to appear at $\mathcal{O}(R^{-5})$. In particular, the difference between droplets and funnels holographic stress energy tensors should only appear at $\mathcal{O}(R^{-5})$. 
\end{subsection}

\end{section}

\begin{section}{Discussion}
To summarise our findings, we have numerically constructed Schwarzschild black droplet solutions suspended over a planar black hole.  These solutions are dual to the ``jammed" phase of a large N strongly coupled CFT.  We find two branches of droplets: long and thin, and that these solutions only exist below a critical temperature $T_\infty/ T_\mrm{BH}\sim0.93$.  We have computed their stress tensor and find generically two power-law regions corresponding to a droplet-like falloff of $R^{-5}$ and a funnel-like falloff of $R^{-1}$.

It would be interesting to study the stability of these droplet solutions.  The short droplet with $T_\infty=0$ were argued to be stable in \cite{Figueras:2011va}.  If they are, then it seems likely that short droplets for small temperature ratios are also stable.   The long droplets, on the other hand, may be unstable to forming a flowing funnel, or perhaps a short droplet.  

If all of our short droplets remain stable, then the critical temperature might be interpreted as a ``melting" or ``freezing" point.  Consider a short droplet at small $T_\infty/T_\mrm{BH}$.  Keeping the boundary black hole fixed, suppose we slowly increase the temperature $T_\infty$.  If we do this slowly enough, the dynamical solution should remain close to the static solution.  Eventually, these static droplets no longer exist, so the system must become fully dynamical, perhaps evolving into a flowing funnel.  The rigid behaviour of the droplet transitions into the more fluid behaviour of a funnel.  

Unfortunately, we cannot directly compare the long and short droplets to each other.  These solutions are not at equilibrium, so their free energy is not well-defined.  One can in principle still compare their entropies and energies.  These quantities are formally infinite, but can be regulated by subtracting the large $R$ behaviour obtained via perturbation theory.   Unfortunately, these quantities are finite only after subtracting down to an $O(R^{-4})$ behaviour, which is beyond our numerical control.  

To complete our understanding of solutions with a Schwarzschild boundary, the flowing funnels need to be constructed.  These solutions would require non-Killing horizons, such as those in \cite{Fischetti:2012ps,Fischetti:2012vt,Emparan:2013fha}.  Additionally, In our solutions, the droplet horizon has the same temperature as the boundary black hole.  It is possible to detune so that these temperatures are not equal \cite{Fischetti:2012vt}.  

In our study, we have focused on boundary black holes that correspond to four-dimensional Schwarzschild.  These boundary black holes do not need to satisfy any field equations, so we are free to choose any metric.  It would be interesting to see what changes as we vary the boundary black hole.  For instance, equilibrium droplets or droplets with $T_\infty/ T_\mrm{BH}>1$ may exist, particularly for boundary black holes that are small relative to their temperature.  

\end{section}
%----- Acknowledgements -------
\vskip 1cm
\centerline{\bf Acknowledgements}
\vskip .5 cm
It is a pleasure to thank Donald Marolf for invaluable discussions. J.E.S.'s work is partially supported by the John Templeton Foundation. B.W. was supported by European Research Council grant no. ERC-2011-StG 279363-HiDGR.

%----- Bibliography ---------
\singlespacing

\bibliography{refs}{}
\bibliographystyle{JHEP}

\end{document}